\documentclass[prb,twocolumn,aps,showpacs,10pt,superscriptaddress]{revtex4}
\usepackage{graphicx,epsf,bm,bbm,color,amsmath,ulem,amssymb,psfrag,float}
\usepackage{comment}

	\newcommand{\beq}{\begin{equation}}
	\newcommand{\be}{\begin{equation}}
	\newcommand{\beqn}{\begin{eqnarray}}
	\newcommand{\eeq}{\end{equation}}
	\newcommand{\ee}{\end{equation}}
	\newcommand{\eeqn}{\end{eqnarray}}

	\newcommand{\ep}{{\epsilon}}

	\newcommand{\uu}{{\vec u}}
\newcommand{\h}{{\vec h}}
	
	\newcommand{\g}{{\delta}}

	\renewcommand{\d}{{\boldsymbol \delta}}
	\renewcommand{\k}{{\boldsymbol k}}
     \newcommand{\R}{{\boldsymbol R}}
        \renewcommand{\r}{{\boldsymbol r}}
	\newcommand{\G}{{\boldsymbol G}}
     \newcommand{\q}{{\boldsymbol q}}
    \newcommand{\w}{{\vec w}}
    \renewcommand{\h}{{\vec h}}
    \newcommand{\s}{{\vec \sigma}}

\newcommand{\bem}{\begin{pmatrix}}
\newcommand{\eem}{\end{pmatrix}}

\newcommand{\red}{\color{red}}

\begin{document}

\title{
Winding vector: how to annihilate two Dirac points with the same charge
}

       \author{Gilles Montambaux}

\address{Laboratoire de Physique des Solides, CNRS, Universit\'e Paris-Sud, Universit\'e Paris-Saclay, F-91405 Orsay, France}

 \author{Lih-King Lim}
\email{lihking@123mail.org}
        \address{Zhejiang  Institute  of  Modern  Physics,  Zhejiang  University,  Hangzhou  310027,  P.  R.  China}
\address{Institute  for  Advanced  Study,  Tsinghua  University,  Beijing  100084,  P.  R.  China}

       \author{Jean-No\"el Fuchs}
       \address{Laboratoire de Physique des Solides, CNRS, Universit\'e Paris-Sud, Universit\'e Paris-Saclay, F-91405 Orsay, France}
\address{Sorbonne Universit\'e, CNRS, Laboratoire de Physique Th\' eorique de la Mati\` ere Condens\' ee, LPTMC, F-75005 Paris, France}

        \author{Fr\'ed\'eric Pi\'echon}
\address{Laboratoire de Physique des Solides, CNRS, Universit\'e Paris-Sud, Universit\'e Paris-Saclay, F-91405 Orsay, France}


\begin{abstract}
  The merging or emergence of a pair of Dirac points may be classified according to whether the winding numbers which characterize them are opposite ($+-$ scenario) or identical ($++$ scenario).
From the touching point between two parabolic bands (one of them can be flat), two Dirac points with the {\it same} winding number emerge under appropriate distortion (interaction, etc), following the $++$ scenario. Under further distortion, these Dirac points  merge following the $+-$ scenario, that is corresponding to  {\it opposite} winding numbers.
This apparent contradiction is solved by the fact that the winding number is actually defined around a unit vector on the Bloch sphere and that this vector rotates during the motion of the Dirac points. This is shown here within the simplest  two-band lattice model (Mielke)  exhibiting a  flat band. We argue on several examples that  the evolution between the two  scenarios is general.
\end{abstract}
\maketitle

{\it Introduction\,--- }
There has been a recent growing interest for various physical systems exhibiting a multiband excitation spectrum with crossing points between the bands. This interest was boosted by the discovery of graphene, where the low energy spectrum is described by a 2D Dirac equation for massless Fermions, giving the name "Dirac point" to such linear crossing point.\cite{graphene}
 In two dimensions, a band touching is a topological defect protected by time-reversal and inversion symmetries.   Such a contact point  is characterized by  a
  winding number $w$ (sometimes confused with a Berry phase\cite{Marzari})  which describes the winding of the phase of the wave function when moving around this point in reciprocal space.  Such singularities  may  emerge or disappear under variation of external parameters under the constraint that the sum of their winding numbers is conserved.\cite{Gail2012,Poincare}

  It has been shown that the merging (or emergence) of {\it two} Dirac points in 2D crystals is described by two ``universal Hamiltonians" depending on the topological properties of the Dirac points that merge.\cite{Gail2012,Poincare} They correspond to  the two scenarios  for winding numbers $(+1, -1) \rightarrow 0$ and $(+1, +1) \rightarrow +2$.
These two Hamiltonians can be written with the help of two Pauli matrices  $\sigma_a, \sigma_b$ ($a,b \in x,y,z$) and three parameters $\Delta, m, c $ or $\Delta, m_a, m_b$:

\be {\cal H}_{+-}= \left(\Delta + {p_x^2 \over 2 m}\right) \,  \sigma_a + c p_y \, \sigma_b \ , \label{Hplusmoins}\ee

\be {\cal H}_{++}= \left(\Delta + {1 \over 2 m_a} (p_x^2  -p_y^2) \right) \,   \sigma_a + { p_x p_y  \over m_b}\, \sigma_b \ . \label{Hplusplus} \ee

For the first Hamiltonian,
the  gapless phase with Dirac points corresponds to $\Delta <0$. When $\Delta  \geq 0$, the Dirac points of {\it opposite} signs ($w=\pm 1$) have merged into a {\it semi-Dirac} spectrum $(\Delta=0$), linear in one direction, {\it quadratic} in the other and then a gap $2 \Delta >0$  opens.\cite{Montambaux:09,Pickett}
The second Hamiltonian describes the  nematic distortion of a quadratic band touching.\cite{Gail2012,Chong,Sun,Gail2011,Dora,Tsai}
A finite value of the parameter $\Delta$ splits this quadratic point into a pair of Dirac points of {\it same charge} along a direction which depends on sgn$(\Delta)$.  The total charge $w=+2$ being conserved, the contact is topologically stable and no gap opens.
  These Hamiltonians are ``universal" in the sense that they provide a unique description of the merging of Dirac points,
 independent of its microscopic realizations. \cite{Tarruell:12,Lim:12,Bellec:13a,Polini:13}
 An additional term proportional to the identity $\sigma_0$    may change the spectrum dramatically but does not change the geometric properties of the wave functions. A quadratic touching point with a flat band enters in the second category with an appropriate $\sigma_0$ term.

\begin{figure}[t!]
\begin{center}
\includegraphics[width=2cm]{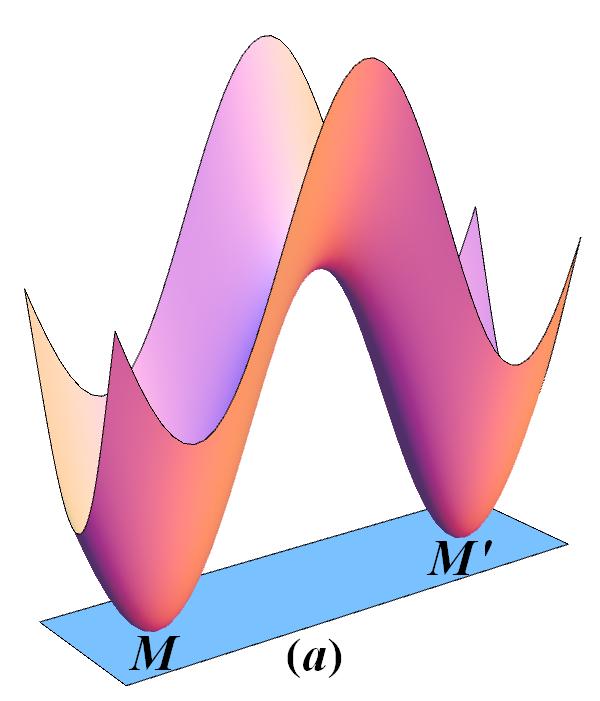} \includegraphics[width=2cm]{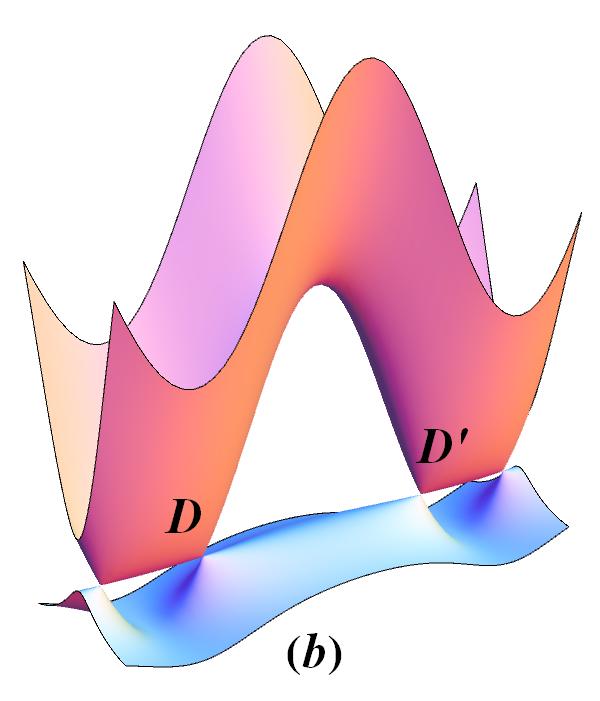}
\includegraphics[width=2cm]{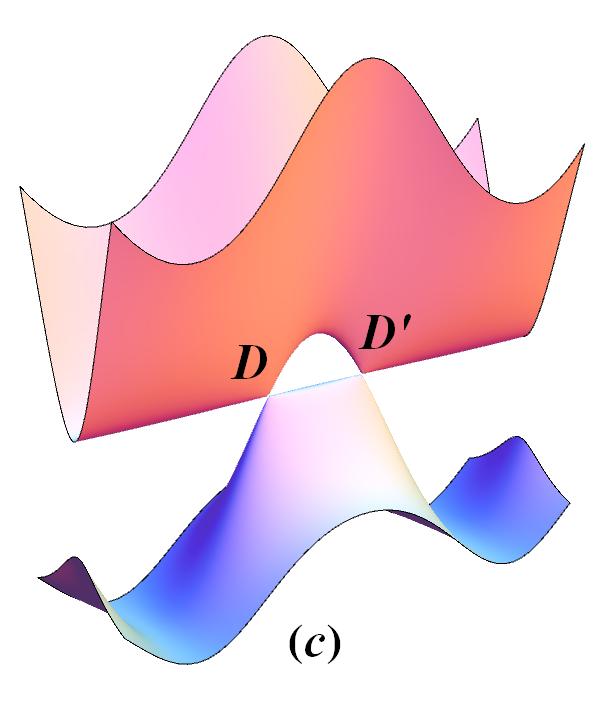} \includegraphics[width=2cm]{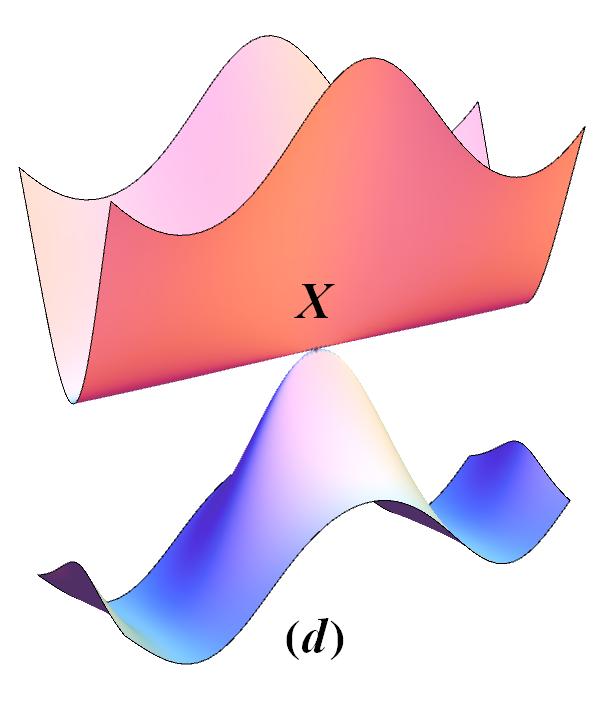}
\vspace{.5 cm}

\includegraphics[width=2cm]{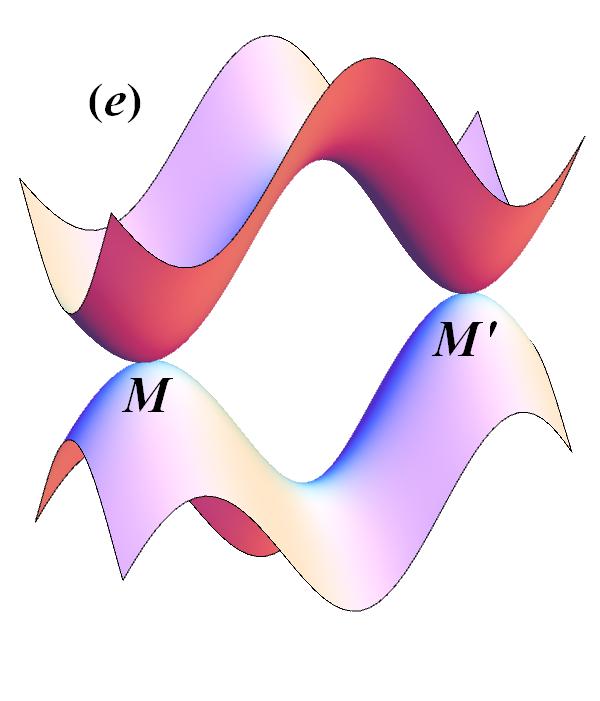} \includegraphics[width=2cm]{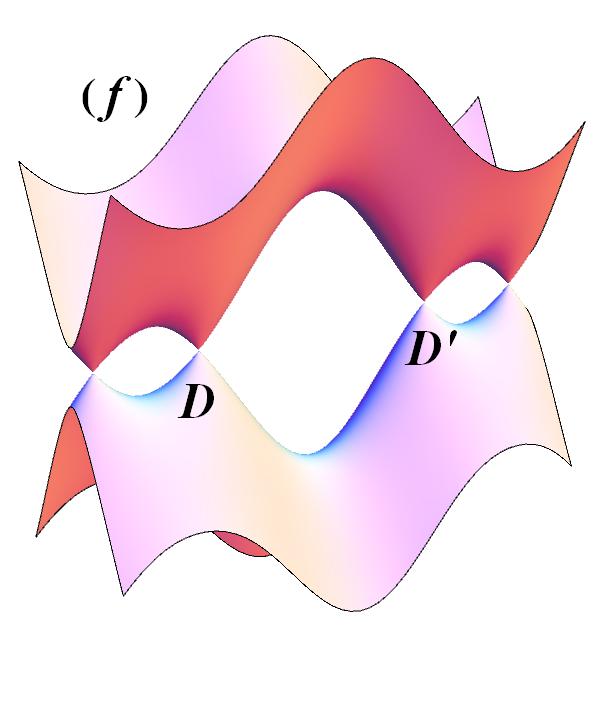}
\includegraphics[width=2cm]{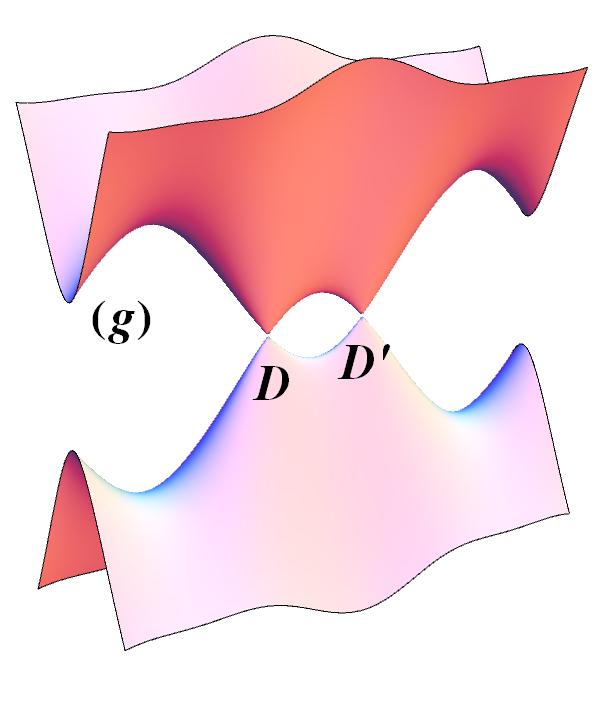}
\includegraphics[width=2cm]{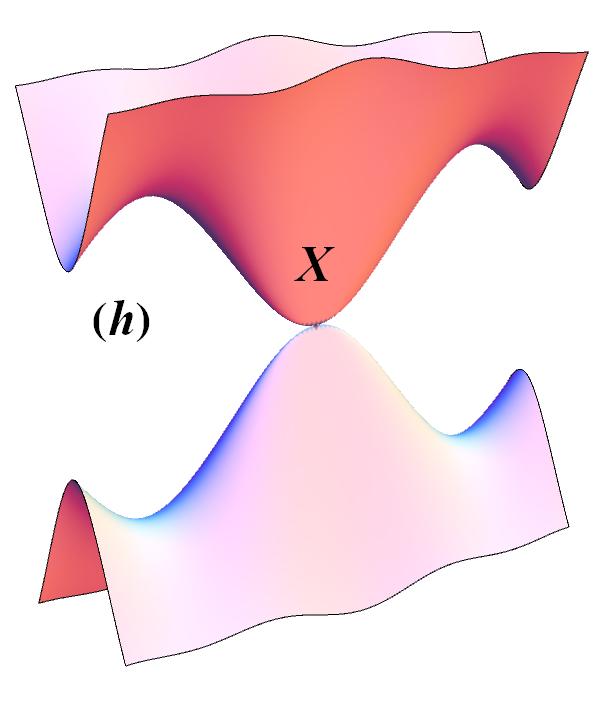}
\end{center}
 \caption{Top (abcd): Emergence and motion of a single pair $(D,D')$ of Dirac points from a flat band touching a quadratic band. The two Dirac points emerging at the M point eventually merge at the X point into an asymmetric semi-Dirac spectrum. Bottom (efgh): Emergence and motion of a pair of Dirac points from a symmetric quadratic spectrum.  The two Dirac points eventually merge into a semi-Dirac spectrum }
 \label{fig:coupes}
\end{figure}

The question that we pose in this letter is the fate of a pair of Dirac points emerging from   a quadratic band crossing, when further distortion is applied.
We find a surprising situation where a unique pair of Dirac points emerges from  a quadratic touching point, and disappears as a semi-Dirac point with gap opening, therefore following  the two different merging scenarios in the  same physical system.
Considering that this pair is unique, its emergence or merging necessary occurs at a time-reversal invariant momentum (TRIM)  $\G/2$ where $\G$ is a reciprocal lattice vector.\cite{remarks}  In the vicinity of these points, the Bloch Hamiltonian takes either either the form
 ${\cal H}_{++}$  with the Pauli matrices  $(\sigma_x, \sigma_z)$, or  the form ${\cal H}_{+-}$ with the matrices  $(\sigma_x, \sigma_y)$ or $(\sigma_z, \sigma_y)$.

This evolution poses a central question: How a pair of Dirac points may emerge or disappear following both scenarios $(++)$ and $(+-)$ while conserving the winding number in  the first Brillouin zone? We show that this is possible by defining a  winding number around  a unit vector which rotates in pseudo-spin space. The resulting winding vectors (see later) of the two Dirac points are parallel at their emergence at the quadratic point, and antiparallel at their merging at the semi-Dirac point.


Here, this evolution is described in the framework of a simple lattice model exhibiting a  contact between a flat band and a quadratic band. This is  a generic model for  systems with  more energy bands, like a deformed Kagome lattice \cite{kagome-lieb} or a honeycomb lattice with  $p_x$-$p_y$ orbitals.\cite{Wu} The latter was probed by a recent experiment   with a polariton lattice of semiconducting micropillars. \cite{c2n1} The deformation of these bands under appropriate lattice distortion  highlights the scenario   described in this letter.\cite{c2n2}
  This experimental work is one of the main motivations for our present study.

     \begin{figure}[h!]
\begin{center}
{\includegraphics[width=8cm]{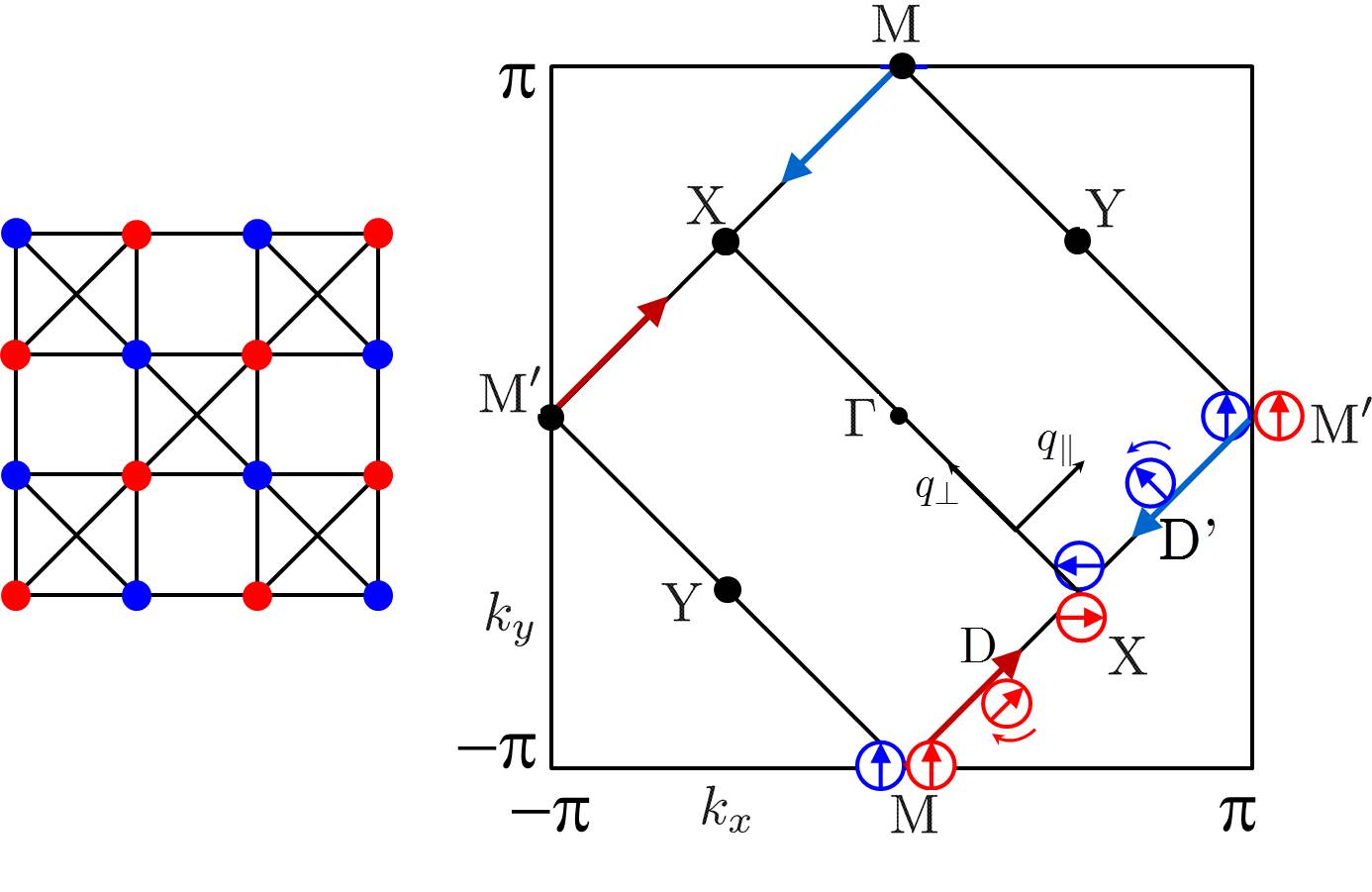}}
\end{center}
 \caption{Left: Mielke model. All bonds have the same hopping parameter $t$. Right: Reciprocal space. The tilted square is the first Brillouin zone. The quadratic touching point is located at the M=M' point from which a pair of Dirac points is created under application of the staggered on-site potential $\delta$.  For $\g >0$, this pair of Dirac points merges at the X point (at the $Y$ point for $\g <0$). The arrows along the edge of the BZ indicate the motion of the Dirac points when the parameter $\delta$ increases  (here $\delta >0$). The boxed arrows indicate the direction of the winding vector $\w$. It rotates from the $y$  to the $x$ direction.}
 \label{fig:mielke}
\end{figure}

\medskip

 {\it Staggered Mielke model\,--- }
In order to study this problem on a concrete simple model, we consider the tight-binding Hamiltonian  visualized in Fig.\ref{fig:mielke}. It has a checkerboard structure will all identical hopping terms $t$. First proposed by Mielke, this is the simplest two-band model exhibiting a flat band. \cite{Mielke}
In addition, we consider the effect of a staggered  on-site potential  $\mp V$ and we set $\g= V / 2 t$.
From the original tight-binding Hamiltonian ${\cal H}$, we introduce the Bloch Hamiltonian
\be {\cal H}(\k)= e^{- i \k . \R}  {\cal H} e^{ i \k . \R} \label{HIBravais} \ee
where $\R$ are the position of the Bravais lattice sites. It has the property $ {\cal H}(\k+\G)= {\cal H}(\k)$ where $\G$ is a reciprocal lattice vector.   Here we define $2t=-1$ and the interatomic distance $a$ is taken as $a=1$. The   Hamiltonian is  written
as
   $ {\cal H}(\k) = \ep_0(\k) \, \sigma_0   + {\cal H}_s(\k) $
 with $\ep_0(\k)=\cos k_x \cos k_y$
and the energy symmetric Hamiltonian:
 \be
  {\cal H}_s(\k)=  \left(
         \begin{array}{cc}
        \g+   \sin k_x \sin k_y &(\cos k_x + \cos k_y )e^{- i k_y}  \\
     (\cos k_x + \cos k_y )e^{ i k_y}   &  - \g -  \sin k_x \sin k_y \\
         \end{array}
       \right)
       \label{Hsmielke}
       \ee

\begin{figure}[t!]
\begin{center}
\includegraphics[width=2.7cm]{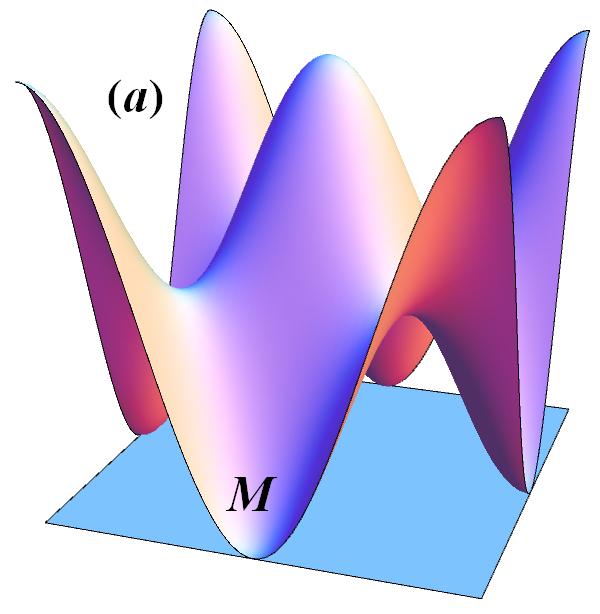} \includegraphics[width=2.7cm]{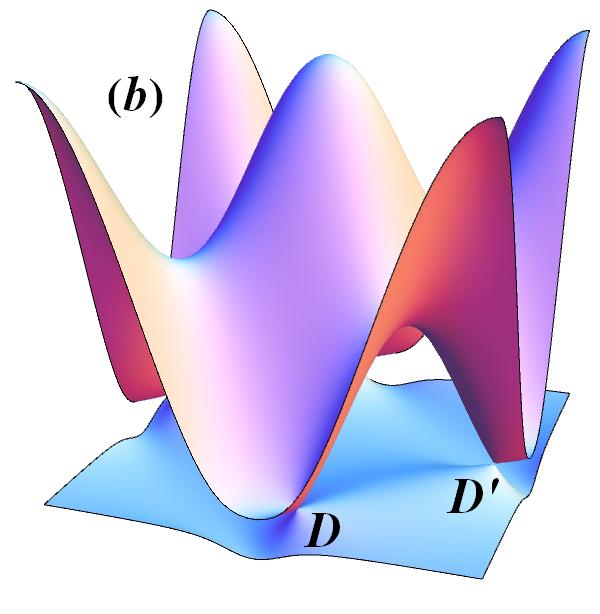}
\includegraphics[width=2.7cm]{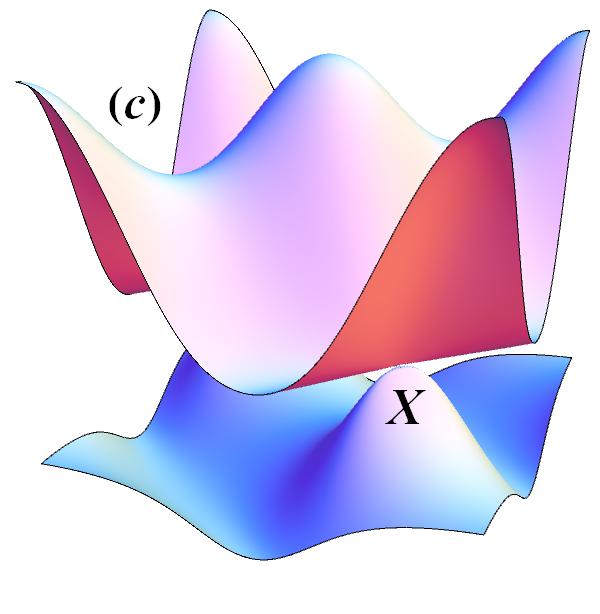}
\end{center}
 \caption{Energy spectrum of the Mielke Hamiltonian ${\cal H}(\k)$ for $\g=0., 0.2,  1$ and $-\pi < k_x, k_y < \pi$.  When $\g$ is finite, two Dirac points appear at the M point and merge at the X point $(\pi/2, -  \pi/2)$ when $\delta=1$. }
 \label{fig:spectres}
\end{figure}
Fig.\ref{fig:spectres} shows  the Mielke spectrum under application of the on-site staggered potential $\delta$
and its   evolution between the two merging   scenarios.
When $\g=0$, the spectrum consists in a dispersive band touching quadratically a flat band of energy $-1$ at the M point $(0,\pi)$  (see Fig. \ref{fig:mielke} for the positions in the reciprocal lattice).
When $\g$ is finite, the flat band becomes dispersive in the energy range  $[-1 - |\g|, -1 + |\g| ]$ and the upper band extends in the range   $[-1 + |\g|, 2 + \sqrt{\g^2 + 4} ]$.  The quadratic touching point splits into two Dirac points  at the energy  $\ep_D= - 1 + |\g|$.  Depending on the sign of $\g$, the Dirac points emerge along one edge or the other of the BZ ($\g>0$ in the figures).
When $\g \rightarrow \pm 1$, these two Dirac points merge at  the   X or Y point $(\pi/2, \mp \pi/2)$   and the spectrum at this merging is  semi-Dirac (here asymmetric), as expected from general arguments discussed below. \cite{Montambaux:09}  For $\g >0$, the full evolution of the spectrum along the merging line (the diagonal M--M', see Fig. \ref{fig:mielke}), is plotted on the top Figs. \ref{fig:coupes}.a-d.
\medskip

Since the geometric properties of this model do not depend on the identity term $\sigma_0$, we concentrate on the symmetric part ${\cal H}_s(\k)$ (\ref{Hsmielke})  of the  Hamiltonian.  Its energy spectrum is symmetric and its evolution along the merging line upon application of the on-site staggered potential $\delta$ is depicted  Figs.\ref{fig:coupes}.e-g. When $\delta=0$, the spectrum is quadratic  and splits into two Dirac points when $\delta$ is finite, very much like the distortion of the quadratic touching in the bilayer graphene spectrum under strain.\cite{Vafek,Gail2011}  Unlike the case of graphene bilayer where there are two quadratic points (in K and K') with opposite winding numbers $w=\pm 2$, here there is  a {\it single} quadratic point in the BZ which occurs at a TRIM (M point). When $\delta \rightarrow 1$  the spectrum converges towards a semi-Dirac point, following the $(+-)$ scenario for  distorted graphene.\cite{Gail2012,Poincare,Montambaux:09,Pickett}


What is the nature of these  Dirac points emerging from a flat band? What are their topological properties?
 To answer these questions, we now concentrate on the motion of the Dirac points along the M--M' line ($k_y = k_x - \pi$), and their merging at the X point.
 \medskip


 {\it Emergence (+,+) at the} M  {\it point\,--- } This situation arises when the parameter $\delta$ is close to $0$.
We expand the symmetric Hamiltonian near the M point located at the south of the BZ (Fig.\ref{fig:mielke}). Writing $\k = (0,-\pi)+ \q$, it has the local form
$ {\cal H}_M=
    {1 \over 2} (q_y^2 - q_x^2) \sigma_x  + (\delta - q_x q_y )\sigma_z  $.
Since we will follow the motion of the Dirac points along the diagonal M--M', it will be convenient to use the rotated coordinates
$q_\parallel = (q_y + q_x)/\sqrt{2}$ and $q_\perp =(q_y - q_x)/\sqrt{2}$
corresponding  to the coordinates respectively along the merging axis and the perpendicular axis.
We find

\be {\cal H}_M=  -q_\parallel  q_\perp \sigma_x +[
  \delta - {1 \over 2} (q_\parallel^2 - q_\perp^2 )]\sigma_z  \ . \label{HM}
\ee

This is the form of universal Hamiltonian ${\cal H}_{++}$ (Eq. \ref{Hplusplus}).  Here unlike the case of bilayer graphene where the low energy Hamiltonian  is written with the Pauli matrices  ($\sigma_x,\sigma_y$), this Hamiltonian is real and involves the matrices ($\sigma_x,\sigma_z$).  When $\g=0$, the associated winding number is $2$ around the $\sigma_y$ direction  and the spectrum is locally quadratic $\ep(\q)= \pm {1 \over 2} (q_\parallel^2 + q_\perp^2)$.
\medskip


 {\it Merging (+,-) at the} X  {\it point\,--- } This situation arises when $\delta$ is close to $1$  (a similar situation occurs at the Y point when $\g$ is close to  $-1$).
Introducing again the coordinates $q_\parallel$ and $q_\perp$ and neglecting higher order terms, the symmetric Hamiltonian in the  vicinity of the X point
has the form

\be {\cal H}_X= \sqrt{2}\,   q_\perp \sigma_y +
[\delta  -1 + {1 \over 2} q_\parallel^2 ]\sigma_z \ ,
\ee
which is the universal Hamiltonian ${\cal H}_{+-}$  (Eq. \ref{Hplusmoins}) written here in $(\sigma_y, \sigma_z)$ space.  It describes the merging of two Dirac points with winding $\pm 1$ around the $\sigma_x$ direction.  When $\g=1$, the two charges annihilate, there is no winding anymore and the spectrum is semi-Dirac $\ep(\q)= \pm \sqrt{2 q_\perp^2 + q_\parallel^4/4}$.
\medskip


     \begin{figure}[h!]
\begin{center}
\includegraphics[width=8cm]{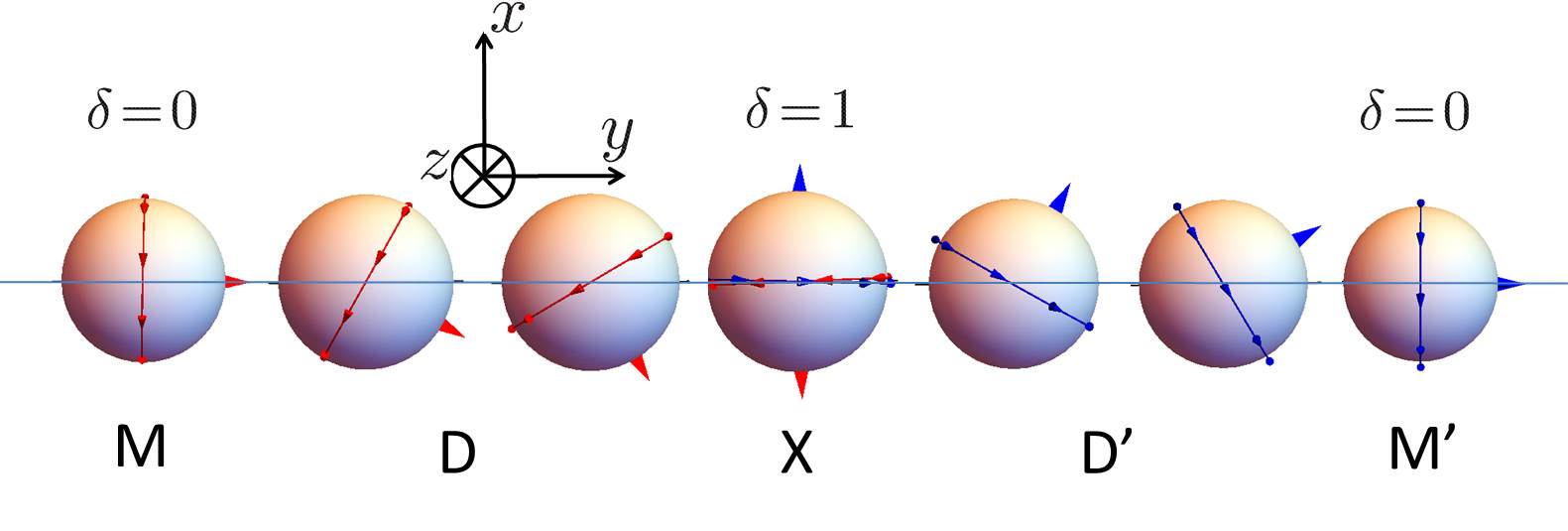}
\end{center}
 \caption{For a given $\delta$, the pseudo magnetic field $\h(\q)$ around a Dirac point rotates along an equator of the Bloch sphere whose orientation varies continuously with $\delta$, from the M (=M') point to the X point. This orientation is characterized by a winding vector (big arrow) which continuously evolves form the $y$ axis to the $x$ axis. }
 \label{fig:motion}
\end{figure}

 {\it Following the moving Dirac points\,--- }
We conclude from these two limits that, upon variation of the parameter  $\delta>0$,  the emergence and merging of Dirac points is described by two universal Hamiltonians,  respectively   ${\cal H}_{++}$ and  ${\cal H}_{+-}$, the first one involving pseudospin components $(\sigma_x,\sigma_z)$ and the other one having components $(\sigma_y,\sigma_z$).  This implies a {\it continuous rotation} in pseudo-spin space during the motion of the Dirac points. To follow this rotation, we now expand locally the Hamiltonian in the vicinity of the two moving Dirac points D, D' of coordinates
$k_x^{D} = \arcsin \sqrt{\delta}$, $k_x^{D'} = \pi - \arcsin \sqrt{\delta}$
and $k_y^{D,D'} = - \pi + k_x^{D,D'}$. We write it in the  form
$ {\cal H}_D= \h \cdot \s   $
with the ``pseudo magnetic field"
\be \h(\q) = v_\parallel q_\parallel \, \uu_z+ v_\perp q_\perp  \,  \uu_\phi \ee
where the  velocities are given by
\be
  v_\parallel = \mp \sqrt{2 \delta(1-\delta)}   \quad , \quad
   v_\perp = \sqrt{2 \delta} \ . \ee
We have introduced the unit vector $\uu_\phi = \cos k_y^D  \uu_x + \sin k_y^D \uu_y $. Therefore for a given value of the parameter $\delta$, the Hamiltonian ${\cal H}_D$ in the vicinity of a Dirac is written is terms of only  {\it two} Pauli matrices $\sigma_z$ and $\sigma_\phi= \s \cdot \uu_\phi$. Around each Dirac point the normalized pseudo magnetic field $\h / |\h|$ rotates along an equator of the Bloch sphere, whose orientation varies when moving the position of the Dirac points from the M point to the X point. Therefore we are led to define a ``{\it winding vector}", perpendicular to this equator, and given by

\begin{align} \w_{D,D'}&= \text{sign}(v_\parallel v_\perp)\,  \uu_z \times \uu_\phi  \nonumber \\
&=  \mp \sqrt{\delta}\,  \uu_x + \sqrt{1 - \delta}\,  \uu_y  \ .\end{align}

\noindent
Fig. \ref{fig:motion} shows the evolution of the winding vector from the M  to the X point. At the M point, two Dirac points emerge from a quadratic touching, with {\it identical} winding vectors $\w= \uu_y$. They merge at the X point with {\it opposite} winding vectors $\pm \uu_x$ (Fig.\ref{fig:motion}).

Near the M point  both velocities vanish before the merging into a quadratic point. Near the X point, the velocity vanishes along the $\parallel$ direction and stays finite along the $\perp$ direction, announcing the merging into a semi-Dirac point.   Finally we note that the motion of Dirac points between the TRIM is  not necessarily along a straight line. It is in general  determined by the vanishing of the off-diagonal element of  ${\cal H}_s({\bf k})$.

\begin{figure}[t!]
\begin{center}
\includegraphics[width=9cm]{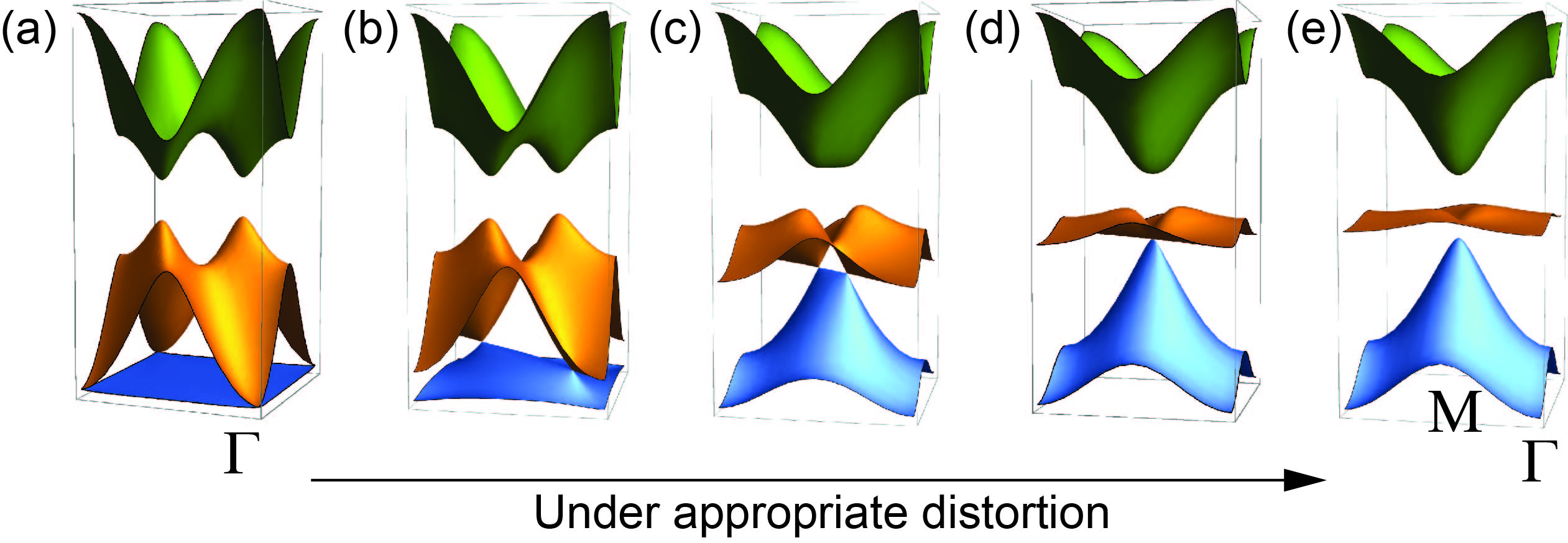}
\end{center}
 \caption{In this 3-band spectrum of a distorted Kagome-like  lattice (with square symmetry), one of the bands is separated from the two others. Under appropriate  variation of hopping parameters, a pair ($++$) of Dirac points emerges at the $\Gamma$ point and disappears as a pair ($+-$) at another TRIM (M). }
 \label{fig:LiebKagome}
\end{figure}

 {\it Multiband systems and experimental motivation\,--- }
For pedagogical purpose, we have chosen to describe in detail a simple two-band problem.  The scenario presented here is generic of more complex situations encountered in multiband spectra that we now briefly illustrate in the  cases of a  3-band and a 4-band problem.
First we consider a square variation of the Kagome lattice which is known to exhibit a flat band touching quadratically a dispersive band.\cite{Huse} Under appropriate variation of hopping parameters, we find that a pair of Dirac points between the two lower bands emerges from the $\Gamma$ point and merges at another TRIM, with a semi-Dirac spectrum preceding the opening of a gap, see Fig. \ref{fig:LiebKagome}.\cite{kagome-lieb}

\begin{figure}[h!]
\begin{center}
\includegraphics[width=8cm]{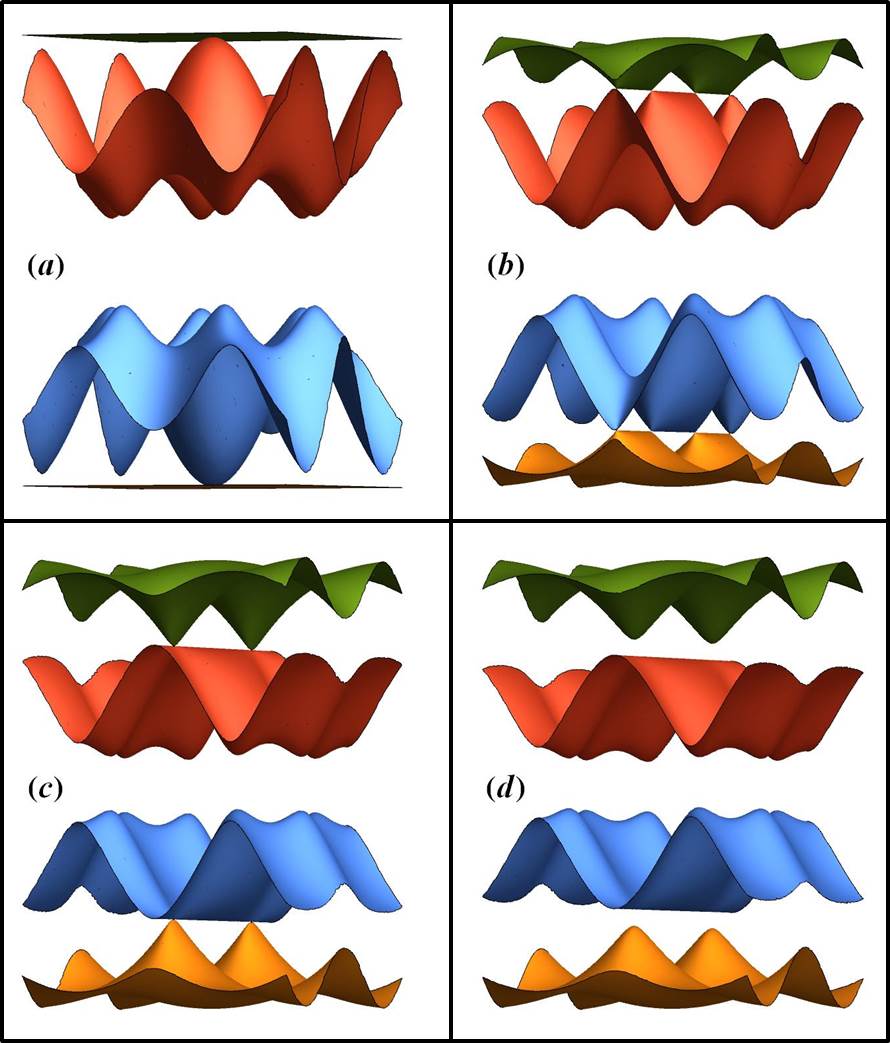}
\end{center}
 \caption{Spectrum of the $p_x$-$p_y$ bands of a graphene-like structure. A staggered potential opens a gap between the lower and upper bands. We exhibit the emergence of a pair of Dirac points (b) from a quadratic point (a), which ultimately merges into a semi-Dirac point (c) preceding the opening of a gap (d).}
 \label{fig:pxpy}
\end{figure}

Then we consider the spectrum of the honeycomb lattice with two orbitals $p_x,p_y$ per site. It  consists in four bands arranged as two bands similar to the $p_z$ bands of graphene sandwiched between two flat bands (Fig. \ref{fig:pxpy}).\cite{Wu}
In addition, a  staggered potential  opens a gap in the middle of the spectrum (like in boron nitride) and separates the two upper bands from the two lower bands. Each  dispersive band touches a flat band at the $\Gamma$ point. We then apply a uniaxial distortion similar to that done in artificial graphenes.\cite{Montambaux:09}  Fig. \ref{fig:pxpy} shows that under this distortion  two Dirac points emerge from               the quadratic touching points (the upper one and the lowest one)  therefore following the $++$ scenario and that they ultimately merge at a TRIM following the $+-$ scenario.

\medskip



 {\it Discussion and perspectives\,--- }
In this letter, we have shown on a simple model, that a Dirac point is characterized by an integer {\it and a direction on the Bloch sphere}, leading to the notion of a winding {\it vector} rather than number. We summarize here the scenario: for two bands, a Bloch Hamiltonian $ {\cal H}(\k)= \h(\k) \cdot  \s $ involves {\it three} Pauli matrices and can be represented as a point on a Bloch sphere, i.e. the position of a normalized pseudo magnetic field $\h/|\h|$. In 2D, contact points between two bands are unstable unless a particular  symmetry protects their existence. In such a case, locally around a contact point $\k_D$, the Hamiltonian $ {\cal H}(\k_D+\q)= h_a(\q)  \sigma_a + h_b(\q) \sigma_b $ involves only {\it two} Pauli matrices $\sigma_a = \uu_a(\k_D) \cdot \s$ and $\sigma_b = \uu_b(\k_D) \cdot \s$.  The pseudo magnetic field is therefore restricted to move on an equator. The contact point is then characterized by the number of times  the latter winds around the direction $\uu_a \times \uu_b$ perpendicular to the equator  when encircling the contact point in reciprocal space.
Most generally the winding vector is given by
\be \w= \left( {1 \over 2 \pi } \oint \vec \nabla \psi \cdot d\q \right) \  \uu_a \times \uu_b\ee
with $\tan \psi= h_b(\q) / h_a(\q)$ and the winding number $|\w|$ is a non-negative integer. For a Dirac point the winding number is 1. At a TRIM, it is 0 or 2, depending on the merging of the two Dirac points. The notion of a winding vector becomes crucial when  its direction changes as the contact point moves in reciprocal space. In the present case, it solves the apparent paradox that a single pair of Dirac points is created with a total winding number of $2$ ($++$ scenario  : $(+1,+1) \rightarrow +2$) and annihilated with a total winding number of $0$ ($+-$ scenario  : $(+1,-1) \rightarrow 0$).

The situation has been studied here within the simplest two-band lattice model. It is universal in the sense that it properly describes the evolution of the crossing point between two bands in a multi-band  system. It describes the structure of the Dirac points emerging from the touching with a flat band. Recent experiments in higher bands of an artificial graphene have successfully shown new pairs of Dirac points emerging from a flat band and whose evolution follows the mechanism described in this letter.\cite{c2n2} Given the universality of this mechanism it should be observed routinely in the many new condensed matter or $2D$ artificial structures exhibiting several bands in the excitation spectrum.
In $3D$, contact points between two bands are generic and do not need any symmetry protection (see e.g. Weyl semi-metals).
There is no winding vector in this case, since these topological defects are characterized by a charge (the  wrapping number), which does not require any direction to be specified.\cite{Volovik}

We acknowledge useful discussions with A. Amo, J. Bloch, A. Mesaros and M. Mili\'cevi\'c. L.-K. L. is supported by Tsinghua University Initiative Research Programme and the 1000 Youth Fellowship China.


\begin{thebibliography}{99}



\bibitem{graphene} For a review on graphene, see A. H. Castro Neto, N. M. R. Peres,
K. S. Novoselov, and A. K. Geim, Rev. Mod. Phys. {\bf 81}, 109 (2009).

\bibitem{Marzari} For a discussion between winding number and Berry phase, see C.-H. Park and N. Marzari, Phys. Rev. B {\bf 84}, 205440 (2011)



  \bibitem{Gail2012} R. de Gail, J.-N. Fuchs, M.O. Goerbig F. Pi\'echon and G. Montambaux, Physica B {\bf 407}, 1948 (2012);
 R. de Gail, M.O. Goerbig and G. Montambaux, Phys. Rev. B {\bf 86}, 045407 (2012)

 \bibitem{Poincare} For a review, see   M.O. Goerbig and G. Montambaux, in {\it Dirac Matter, Poincar\'e Seminar}, Prog. Math. Phys. {\bf 71}, 25 (2016)



\bibitem{Montambaux:09} G. Montambaux, F. Pi\'echon, J.-N. Fuchs and M.O. Goerbig, Phys. Rev. B {\bf 80}, 153412 (2009); Eur. Phys. J. B  {\bf 72}, 509 (2009)

     \bibitem{Pickett}V. Pardo and W.
E. Pickett, Phys. Rev. Lett.
{\bf 102},
166803 (2009)



\bibitem{Chong} Y.D. Chong, X.-G. Wen and M. Solja\u{c}i\'c, Phys. Rev. B {\bf 77}, 235125 (2008)

\bibitem{Sun} K. Sun, H. Yao, E. Fradkin and  S. A. Kivelson,
 Phys. Rev. Lett. {\bf 103}, 046811 (2009).

\bibitem{Gail2011} R. de Gail, M.O. Goerbig, F. Guinea, G. Montambaux, and A.H. Castro Neto, Phys. Rev. B {\bf 84}, 045436 (2011)

 \bibitem{Dora} B. Dora, I.F. Herbut and R. Moessner,  	Phys. Rev. B {\bf  90}, 045310 (2014)

 \bibitem{Tsai} W.-F. Tsai, C. Fang, H. Yao and J. Hu, New. J. Phys. {\bf 17}, 055016 (2015)



\bibitem{Tarruell:12} L. Tarruell, D. Greif, T. Uehlinger, G. Jotzu, and T. Esslinger, Nature {\bf 483}, 302 (2012).

\bibitem{Lim:12} L.-K. Lim, J.-N. Fuchs and G. Montambaux,
Phys. Rev. Lett. {\bf 108}, 175303 (2012)


\bibitem{Bellec:13a} M. Bellec, U. Kuhl, G. Montambaux, and F. Mortessagne, Phys. Rev. Lett. {\bf 110}, 033902 (2013).


\bibitem{Polini:13} M. Polini, F. Guinea, M. Lewenstein, H. C. Manoharan and V. Pellegrini,
  Nature Nanotech. {\bf 8}, 625 (2013)

  \bibitem{remarks}   Quadratic band crossings with $w=2$ necessarily occur at the  M point or at the $\Gamma$  point (due to $C_4$ or $C_6$ symmetries).\cite{Chong,Sun} We do not consider here a quadratic band touching with $w=0$, which necessarily splits into at least four Dirac points. A  semi-Dirac band crossing may occur at any of the four points $\Gamma$, M , X or Y (or $\Gamma$, M$_1$, M$_2$, M$_3$ for the triangular symmetry).



\bibitem{kagome-lieb} J.-N. Fuchs, L.-K. Lim, F. Pi\'echon and G. Montambaux, in preparation


\bibitem{Wu} C. Wu, D. Bergman, L. Balents, and S. Das Sarma, Phys.
Rev. Lett. {\bf 99}, 070401 (2007)

\bibitem{c2n1}
M. Mili\'cevi\'c, T. Ozawa, G. Montambaux, I. Carusotto, E. Galopin, A. Lema\^itre, L. Le Gratiet, I. Sagnes, J. Bloch, and A. Amo
Phys. Rev. Lett. {\bf 118}, 107403 (2017)



\bibitem{c2n2} M. Mili\'cevi\'c,  {\it at al.}, in preparation


\bibitem{Mielke} A. Mielke, J. Phys. A {\bf 24}, 3311 (1991)

\bibitem{Vafek} O. Vafek and K. Yang, Phys. Rev. B {\bf 81}, 041401  (2010)





\bibitem{Huse} Y. Xiao, V. Pelletier, P. M. Chaikin and D. A. Huse, Phys. Rev. B  {\bf 67}, 104505 (2003)





\bibitem{Volovik} G. Volovik, {\it The Universe in a Helium Droplet},
 Oxford university press (2009)




\end{thebibliography}
\end{document}